\documentclass[
    ,final            
  ]
  {aipproc}

\usepackage{amsmath,amssymb}

\layoutstyle{8x11single}

\newcommand{\cd}{\! \cdot \!}
\newcommand{\be}{\begin{equation}}
\newcommand{\ee}{\end{equation}}
\newcommand{\ba}{\begin{eqnarray}}
\newcommand{\ea}{\end{eqnarray}}

\begin{document}

\title{Transport coefficients in superfluid neutron stars}


\author{Laura Tolos} { 
  address={Instituto de Ciencias del Espacio (IEEC/CSIC) Campus Universitat Aut\`onoma de Barcelona, Facultat de Ci\`encies, Torre C5, E-08193 Bellaterra (Barcelona), Spain}, 
   altaddress={Frankfurt Institute for Advances Studies. Johann Wolfgang Goethe University, Ruth-Moufang-Str. 1,
60438 Frankfurt am Main, Germany}  
}

\author{Cristina Manuel}{
  address={Instituto de Ciencias del Espacio (IEEC/CSIC) Campus Universitat Aut\`onoma de Barcelona, Facultat de Ci\`encies, Torre C5, E-08193 Bellaterra (Barcelona), Spain}
}

\author{Sreemoyee Sarkar}{
  address={Tata Institute of Fundamental Research, Homi Bhaba Road, Mumbai-400005, India} 
}

\author{Jaume Tarrus}{
address={Physik Department, Technische Universit\"at M\"unchen, D-85748 Garching, Germany} 
}

\begin{abstract}
We study the shear and bulk viscosity coefficients as well as the thermal conductivity as arising from the collisions among phonons in superfluid neutron stars. We use effective field theory techniques to extract the allowed phonon collisional processes, written as a function of the equation of state  and the gap of the system.  The shear viscosity  due to phonon scattering is compared to calculations of that coming from electron collisions. We also comment on the possible consequences for r-mode damping in superfluid neutron stars. Moreover, we find that phonon collisions give the leading contribution to the bulk viscosities in the core of the neutron stars. We finally obtain a temperature-independent thermal conductivity from phonon collisions and compare it with the electron-muon thermal conductivity in superfluid neutron stars.
\end{abstract}

\maketitle

\section{Introduction}

The study of superfluidity in neutron stars has been a matter of interest over decades
after Migdal's observation that superfluidity may occur in the core of compact stars \cite{Migdal}. 
Different neutron star phenomena such as pulsar glitches, star cooling or the dynamics of the neutron star's oscillations might be explained assuming superfluidity inside neutron stars.

Some of these phenomena are  determined by the behavior of transport coefficients in the presence of superfluidity.  The r-modes, which are toroidal modes which only occur in rotating stars, with the Coriolis force acting as their restoring force, are generically unstable in all rotating stars through their coupling to  gravitational radiation (GR) emission~\cite{Andersson:2000mf,Lindblom:2001}.
When dissipative phenomena, such as shear and bulk viscosities, damp these r-modes the star can rotate without losing  angular momentum to GR. Thus, as most of the known neutron stars are pulsars and their rotation frequencies are known with a lot of accuracy, the study of r-modes can be used to constrain the internal stellar structure and the possibility of superfluidity. Also, the cooling of a neutron star could give some hints on its composition and the presence of superfluidity  since it depends on the rate of neutrino emission, the photon emission rate, the specific heat and  in young stars on the thermal conductivity \cite{Pethick:1991mk,Yakovlev:2004iq}. 

At low temperatures neutron matter superfluidity  occurs after the appearance of a quantum condensate, associated to neutron pairing. The condensate spontaneously breaks the global $U(1)$ symmetry associated to baryon number conservation. Thus, the Goldstone theorem predicts the existence of a low energy mode that explains the property of superfluidity, the so-called superfluid phonon.

In this manuscript we explore the phonon contribution to the shear \cite{Manuel:2011ed,Manuel:2012rd} and bulk viscosities \cite{Manuel:2013bwa} as well as to the thermal conductivity  \cite{Manuel:2014kqa} for superfluid neutron matter in the core of neutron stars. In order to compute the phonon contribution  one needs to assess the relevant phonon collisions which are responsible for the transport phenomena.  In this paper we exploit the universal character of effective field theories (EFT)  to present a very general formulation  to determine the leading phonon interactions \cite{Son:2002zn,Son:2005rv}, that  depends on the equation-of-state (EoS) and the superfluid gap of the system.  

\section{Phonon scattering in superfluid neutron matter}

The superfluid phonon is the Goldstone mode associated to the spontaneous symmetry breaking
of a $U(1)$ symmetry, which corresponds to particle number conservation. EFT techniques can be used to write down the effective Lagrangian associated to the superfluid phonon. The effective Lagrangian is expressed as an expansion in derivatives of the Goldstone field, the terms of this expansion being restricted by symmetry considerations. The coefficients of the Lagrangian can be computed from the microscopic theory, through a standard matching procedure. In particular, for the phonons the EoS fixes the leading-order (LO) effective Lagrangian.

At the lowest order in a derivative expansion, the Lagrangian reads \cite{Son:2002zn,Son:2005rv}
\begin{eqnarray}
\label{LO-Lagran}
\mathcal{L}_{\rm LO} &=&P (X) \ , \nonumber \\
 X &=& \mu-\partial_t\varphi-\frac{({\bf \nabla}\varphi)^2}{2m} \ ,
\end{eqnarray}
where $P(\mu)$ and $\mu$ are the pressure and chemical potential, respectively, of the superfluid at $T=0$. The quantity
$\varphi$ is the phonon field and $m$ is the mass of the
particles that condense. After a Legendre
transform, one can get the associated Hamiltonian, which has the same form as the one used by Landau to obtain
the self-interactions of the phonons of $^4$He \cite{Son:2005rv,IntroSupe}.

After a Taylor expansion of the pressure, and rescaling of the phonon field  ($\phi$) to have a canonically normalized kinetic term, the Lagrangian for the phonon field is given by
\begin{eqnarray}
\label{comlag}
\mathcal{L}_{\rm LO}&&=\frac{1}{2}\left((\partial_t\phi)^2-v^2_{\rm ph}({\bf \nabla}\phi)^2\right) \nonumber \\
&&-g\left((\partial_t \phi)^3-3\eta_g \,\partial_t \phi({\bf \nabla}\phi)^2 \right)
+\lambda\left((\partial_t\phi)^4-\eta_{\lambda,1} (\partial_t\phi)^2({\bf \nabla}\phi)^2+\eta_{\lambda, 2}({\bf \nabla}\phi)^4\right)
+ \cdots
\end{eqnarray}
The different  phonon self-couplings of Eq.~(\ref{comlag}) can be  expressed
as different ratios of derivatives of the pressure with respect to the chemical potential \cite{Escobedo:2010uv}, or, similarly, in terms of the speed of sound at $T=0$ and derivatives
of the speed of sound with respect to the mass density. The speed of sound at $T=0$ is given by 
\begin{equation}
\label{phspeed}
v_{\rm ph}=   \sqrt{\frac{\partial P}{\partial {\rho}} } \equiv c_s \ ,
\end{equation}
where ${\rho}$ is the mass density.

The three and four phonon self-coupling constants can be expressed as 
\be
\begin{split}
&g=\frac{1-2 u}{6 c_s \sqrt{\rho}}\,,\qquad  \eta_g=\frac{c^2_s}{1-2 u}\,,\qquad \lambda=\frac{1-2 u(4-5u)-2 w \rho}{24c^2_s\rho}\,,\\
&\eta_{\lambda\,,1}=\frac{6c^2_s(1-2 u)}{1-2u(4-5 u)-2w\rho}\,,\qquad  \eta_{\lambda\,,2}=\frac{3c^4_s}{1-2u(4-5 u)-2w\rho} \ ,
\label{eq:relations}
\end{split}
\ee
with
\be
u=\frac{\rho}{c_s}\frac{\partial c_s}{\partial\rho}\,, \quad w=\frac{\rho}{c_s}\frac{\partial^2 c_s}{\partial\rho^2}\,.
\label{precoup}
\ee
The  dispersion law obtained from this Lagrangian at tree level is exactly  $E_p = c_s p $.  

A next--to--leading order (NLO)  Lagrangian in a derivative expansion can be constructed as well (see, for example, the expression for cold Fermi gas in the unitarity limit \cite{Son:2005rv}). For our purposes,  we will simply compute the leading order $T$ corrections of the transport coefficients so we do not need it. It is however relevant that  the phonon dispersion law suffers corrections when one goes beyond the LO expansion. In fact, at NLO the phonon dispersion law  reads
\be
\label{NLOdisp-law}
E_P = c_s p ( 1 + \gamma p^2) \ .
\ee
The sign of $\gamma$ determines whether the decay of one phonon into two is kinematically allowed or not. Only dispersion laws that curve upward can allow such processes. 

The value of $\gamma$ can be computed through a matching procedure with the underlying microscopic theory. For neutrons paring  in a $^1S_0$ channel within neutron stars it can be seen that  \cite{Manuel:2014kqa}
 \be
\gamma=-\ \frac{v_F^2}{45 \Delta^2} , 
\ee
with $v_F$ being the Fermi velocity and $\Delta$ the value of the gap in the $^1S_0$ phase. We will assume that $\gamma$ takes this same value in the $^3P_2$ phase, with $\Delta$ being the angular averaged value of the gap in that phase.  Thus, considering that $\gamma <0$  the first allowed phonon scattering will be binary collisions.  Explicit values for the gap function used in this work are given in the next section.

\section{Equation of state and gap of superfluid neutron matter}
\label{EOS-section}

The EoS for neutron matter in neutron stars is the key ingredient to calculate the speed of sound at $T=0$ as well as the different phonon self-couplings.  A common benchmark for a nucleonic EoS is the one  by  Akmal,  Pandharipande and  Ravenhall (APR) \cite{ak-pan-rav} in $\beta$-stable nuclear matter. Heiselberg and Hjorth-Jensen  parametrized the APR EoS of nuclear matter in a simple form \cite{hei-hjo}, which will subsequently  be used in this paper. The effect of neutron pairing $\Delta$ in the EoS is not considered  because of $\Delta/\mu << 1$.

\begin{figure}[t]
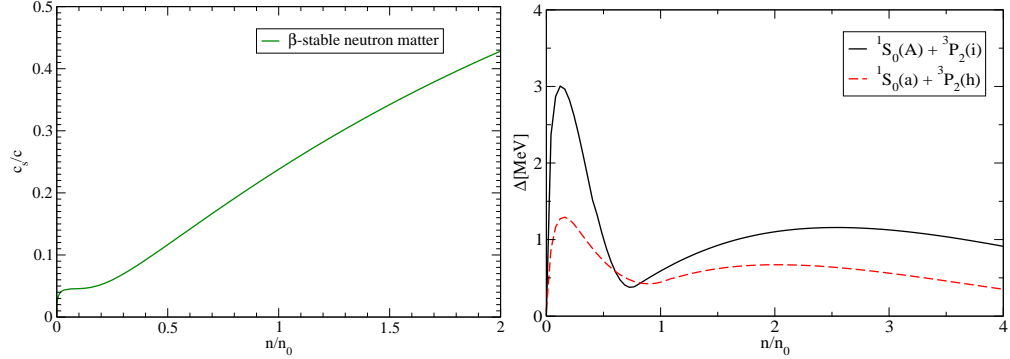

\includegraphics[width=0.4\textwidth]{cs_aip.eps}
\includegraphics[width=0.4\textwidth,height=4.7cm]{gap_aip.eps}
\caption{Left: Speed of sound $c_s$ with respect to the speed of light  $c$  for $\beta$-stable nuclear matter as a function of density. Right: The sum of the $^1S_0$ and angle-averaged $^3P_2$ neutron gaps as a function of density  for  two different gap models.}
\label{speeds-gap}
\end{figure}

The  ratio of the speed of sound $c_s$ with respect to the speed of light $c$ is shown on the l.h.s of Fig.~\ref{speeds-gap}  for  $\beta$-stable nuclear matter as a function of the density. We observe that the assumption that the APR model for $\beta$-stable nuclear matter is non-relativistic breaks down at densities of the order of 1.5-2 $n_0$. For those densities, relativistic effects appear as the APR EoS includes not only two nucleon but also three nucleon interactions.

On  the r.h.s of Fig.~\ref{speeds-gap} we present two very different gap models that are used in this work as a function of the density in order to illustrate the model dependence of our results. The first model, named hereafter $^1S_0(A)$$+$$^3P_2(i)$,  consists of the $^1S_0$ neutron gap that results from the BCS approach using different bare nucleon--nucleon interactions that converge towards a maximum gap of about 3 MeV at $p_F \approx 0.85 {\rm fm}^{-1}$ (parametrization $A$ of Table I in Ref.~\cite{Andersson:2004aa}). The anisotropic $^3P_2$ neutron gap is more challenging and not fully understood as one must extend BCS theory and calculate several coupled equations while including relativistic effects since the gap extends for densities inside the core. We have taken the parametrization $i$ (strong neutron superfluidity in the core) of Table I in Ref.~\cite{Andersson:2004aa} for the $^3P_2$ neutron angular averaged value, which presents a maximum value for the gap of approximately 1 MeV. The second model considered, $^1S_0(a)$$+$$^3P_2(h)$, goes beyond BCS for the $^1S_0$ neutron gap as it incorporates medium polarization effects (parametrization $a$). The maximum value for the gap is then reduced to 1 MeV.  Moreover, for the $^3P_2$ neutron gap we have taken into account the parametrization $h$ (strong neutron superfluidity) with a maximum value of about 0.5 MeV. The corresponding transition temperatures from the superfluid to the normal phase for both models are $T_c \sim 1/2 \Delta \gtrsim 0.25 \times 10^8 K$.

\section{Shear viscosity and the r-mode instability window}

The shear viscosity $\eta$ is as a dissipative term in the energy-momentum tensor $T_{ij}$. For small deviations from equilibrium one
finds that
\be
\label{shear-stress}
 \delta T_{ij}=- \eta \tilde V_{ij}  \equiv  - \eta\left( \partial_i V_j+ \partial_j V_i -\frac 23 \delta_{ij} \nabla \cd {\bf V} \right) \ , 
  \ee
where ${\bf V}$ is the three fluid  velocity of the normal component of the system.

The superfluid phonon contribution to  the energy-momentum tensor
of the system is given by
 \be
 T_{ij}= c_s^2 \int \frac{d^3 p}{(2 \pi)^3}  \frac{ p_i p_j}{E_p} f(p,x)  \ , 
  \ee
where $f$ is the phonon distribution function. The distribution function obeys the Boltzmann equation \cite{IntroSupe}
 \be
  \label{transport}
   \frac{df}{dt} = \frac{\partial
f}{\partial t}+ \frac{\partial E_p}{\partial \bf p} \cdot \nabla f= C[f] \ ,
\ee
being in the superfluid rest frame, and
$C[f]$ is the collision term. For the computation of the collision term it is enough to consider binary collisions, as previously discussed.

In order to compute the shear viscosity we proceed by considering small departures from equilibrium to the phonon distribution function and linearizing the corresponding transport equation \cite{Manuel:2011ed}. It is then necessary to use variational methods in order to solve the transport equation, as in Refs.~\cite{Manuel:2004iv,Alford:2009jm,Rupak:2007vp}. The final expression for the shear viscosity is given by \cite{Manuel:2011ed}
\be
\eta=\left( \frac{2 \pi}{15} \right)^4 \frac{T^8}{c_s^8} \frac{1}{M} \ ,
\ee  
where $M$  is a multidimensional integral that contains the thermally weighted scattering matrix for phonons.

\begin{figure}[t]
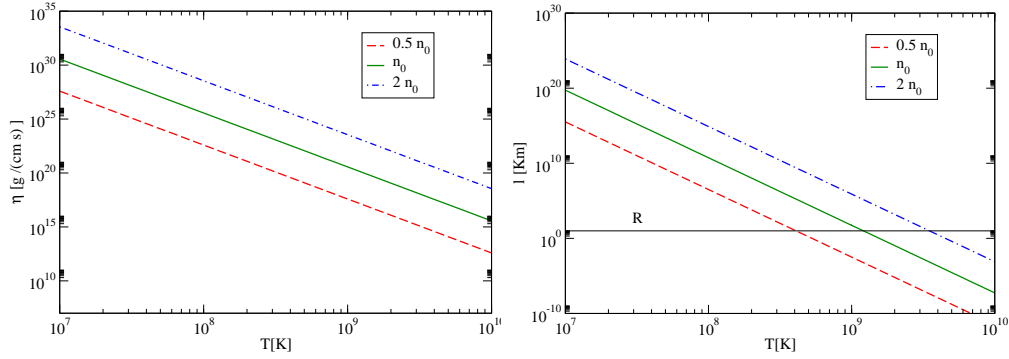

\includegraphics[width=0.4\textwidth]{shear_dens_temp_aip.eps}
\includegraphics[width=0.4\textwidth]{mean_free_path_aip.eps}
\caption{ Left: Shear viscosity for different densites as a function of temperature in $\beta$-stable neutron matter. Right: Phonon mfp in $\beta$-stable matter as function of temperature for different densities. The straight line indicates the radius of the neutron star $R = 10 \ {\rm Km}$.}
\label{fig:shear}
\end{figure}

On l.h.s. of Fig.~\ref{fig:shear} we show that the shear viscosity due to the binary collisions of superfluid phonons scales with temperature
as  $1/T^5$. This indeed results from a very simple dimensional analysis and it is a universal feature that occurs in other superfluid systems  such as ${\rm^4 He}$ \cite{IntroSupe} or superfluid cold atoms at unitary \cite{Rupak:2007vp,Mannarelli:2012su}. The choice of EoS  determines the exact numerical values of  $\eta$.

A relevant discussion is to know the temperature regime  where our results are applicable. Hydrodynamics is only valid when the mean-free path (mfp) is smaller that the typical macroscopic length of the system, in this case, the radius of the star. We show on the r.h.s. of Fig.~\ref{fig:shear} the mfp of phonons for different densities as a function of temperature. We also indicate the limit of 10 Km for the radius of the star.  The mfp $l$ was extracted from the computation of $\eta$ in Ref.~\cite{Alford:2009jm}
\ba
l=\frac{\eta}{n <p>} ,
\label{eq1}
\ea
where $<p>$ is the thermal average momentum, and $n$ is the phonon density:
\ba
<p>=2.7 \frac{T}{c_s} \ , \qquad
n=\int \frac{d^3p}{(2\pi)^3} f_p=\xi(3) \frac{T^3}{\pi^2 c_s^3} .
\label{eq2}
\ea

We observe that for  temperatures below $T\sim 10^9$ K, the
phonon mfp is bigger than the size of the star, thus indicating that
a hydrodynamical description of the phonons of the star is questionable below this temperature.
We note that the critical temperature for the phase transition to the normal phase is $T_c \sim 10^{10}$ K.

\subsection{r-mode instability window}

R-mode oscillations have been studied extensively in the literature and various
damping mechanisms have been proposed~\cite{Andersson:2000mf}. When viscous damping of the r-mode is taken into account the star is stable at low frequencies, or
at very low or high temperatures, but there is typically an instability region at enough high frequencies~\cite{Lindblom:1998wf, Andersson:1998ze}. 
Since the spin rate of several neutron stars falls in this instability window, most of the studies of
the r-modes are focused in modifying or eliminating the instability window. 

In order to assess the r-mode instability window one has to compute the value of the different time scales associated to  
both the instability due to gravitational wave emission $\tau_{\rm GR}$ , and to the various dissipative processes that may damp the r-mode in the star. In this work we only analyze the effect of the shear viscosity, $\tau_{\eta}$, in the r-mode instability window by solving
\be
- \frac{1} { |\tau_{\rm GR} (\Omega) |} +\frac{1}{\tau_{\eta} (T)} = 0  \ ,
\ee
 where the characteristic time scales are given by the expressions of Ref.~\cite{Lindblom:1998wf} (in C.G.S. units)
 \be
\frac{1} { |\tau_{\rm GR} (\Omega) |} = \frac{32 \,\pi \,G\, \Omega^{2l+2}}{c^{2l+3}} \frac{(l-1)^{2l}}{\left( (2l+1)!! \right)^2} \left( \frac{l+2}{l+1} \right)^{2 l+2}  \int^R_0 \rho r^{2 l+2} dr \ ,
\ee  
and 
\be
\frac{1}{\tau_{\eta} (T)}  = (l-1) (2l +1) \int^R_{R_c} \eta r^{2l} dr   \left(\int^R_0 \rho r^{2l+2} dr \right)^{-1}  \ .
\label{ec-hidro-shear}
\ee
We study only  r-modes with  $l = 2$ because they are the dominant ones \cite{Andersson:2000mf,Lindblom:1998wf}.  It has been seen in the literature that electron collisions are one of the most efficient damping mechanisms in neutron stars \cite{Shternin:2008es}. Thus, we consider both the shear viscosity as arising from electrons and phonons. For the calculation of the characteristic time associated to the phonon shear viscosity we only take into account the phonon contribution in the hydrodynamical regime. This is achieved by introducing a temperature-dependent lower limit ($R_c$) in the integral for the shear viscosity in Eq.~(\ref{ec-hidro-shear}).

\begin{figure}[t]
\includegraphics[width=0.55\textwidth,height=6cm]{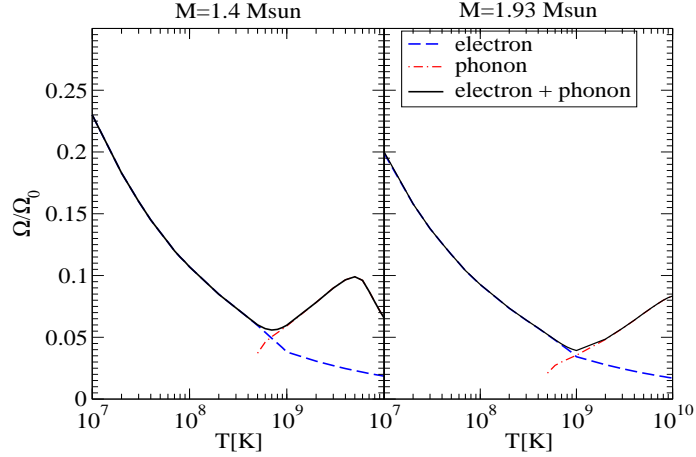}
\caption{ R-mode instability window for superfluid neutron stars. The critical
(normalized) frequency is plotted for two neutron star mass configurations (1.4 ${\rm M_{sun}}$ and 1.93 ${\rm M_{sun}}$) as a function of temperature for the different dissipative processes studied. Here $\Omega_0 =\sqrt{G \pi \bar{\rho}}$, where $\bar{\rho}$ is the average mass density of the star.}
\label{fig:freq}
\end{figure}

In Fig.~\ref{fig:freq} we show the r-mode instability  window of the neutron star as arising from considering shear viscous damping
for two star mass configurations, 1.4 ${\rm M_{sun}}$ and 1.93 ${\rm M_{sun}}$.  The critical frequency in Fig.~\ref{fig:freq} is given in terms of $\Omega_0 =\sqrt{G \pi \bar{\rho}}$, where $\bar{\rho}$ is the average mass density of the star, and it is  displayed as a function of the temperature for the different processes studied. We  display the estimate from Shternin and Yakovlev that considers  the longitudinal and transverse plasmon exchange in superconducting matter with a transition temperature of $T_{cp} \sim 10^9 {\rm K}$ \cite{Shternin:2008es}  (dashed lines) and the hydrodynamical phonon shear viscosity  (dashed-dotted lines). Moreover, we display the sum of both the shear viscosity due to electron collisions from Shternin and Yakovlev and the hydrodynamical  phonon shear viscosity (solid lines). We observe that for low temperatures  the electron contribution to the shear viscosity dominates and governs the processes of r-mode damping. The dissipation stemming from phonon processes in the hydrodynamical regime is non-negligible for $T  \gtrsim7 \times 10^8 {\rm K}$ for a star of 1.4 ${\rm M_{sun}}$ while temperatures of $T \gtrsim10^9 {\rm K}$ are needed for a mass configuration of 1.93 ${\rm M_{sun}}$. 

\section{Bulk viscosities}
The bulk viscosity ($\zeta$) enters as coefficient in  the dissipative hydrodynamic equations.  In superfluid matter there are four bulk viscosity coefficients \cite{IntroSupe}. While the physical meaning of $\zeta_2$ is the same as in a normal fluid, $\zeta_1, \zeta_3, \zeta_4$ refer to dissipative processes which lead to entropy production only in the presence of a space--time dependent relative motion between the superfluid and normal fluid components. 

One can extract the values of the bulk viscosities following a method developed by Khalatnikov \cite{IntroSupe}, which consists of studying the dynamical evolution of the phonon number density. It has been shown in Ref.~\cite{Escobedo:2009bh} that this method is equivalent to the method of computing the bulk viscosities using a Boltzmann equation for the phonons in the relaxation time approximation.  For small departures from equilibrium and for small values of the normal and superfluid velocities it turns out that~\cite{IntroSupe}
\be
\zeta_i = \frac{T}{\Gamma_{ph}} \, C_i \ , \qquad i=1,2,3,4 \ ,
\label{bulkstatic}
\ee 
with $\Gamma_{ph}$ the phonon decay rate and 
\be
C_1 = C_4 = -I_1 I_2 \ , \qquad  C_2 = I_2^2 \ , \qquad  C_3 = I_1^2  \ .
\label{coef}
\ee
The quantities $I_1$ and $I_2$ are defined as 
\ba
I_1&=&\frac{60 T^5}{7c^7_s \pi^2}\left(\pi^2\zeta(3)-7\zeta(5)\right)\left(c_s\frac{\partial B}{\partial \rho}-B\frac{\partial c_s}{\partial \rho }\right)\,, \nonumber \\
I_2&=&-\frac{20 T^5}{7c^7_s \pi^2}\left(\pi^2\zeta(3)-7\zeta(5)\right)\left(2Bc_s+3\rho  \left(c_s\frac{\partial B}{\partial \rho}-B\frac{\partial c_s}{\partial \rho }\right)\right)\,,
\label{i1-i2}
\ea
where  $B=c_s \gamma$ and $\zeta(n)$ is the Riemann zeta function. Note that in order to have non--vanishing values of the bulk viscosities one needs to consider the phonon dispersion law beyond linear order.

For astrophysical applications it is however more important to compute the bulk viscosity coefficients when the perturbation that leads the system out  of equilibrium is periodic in time.
It is then easy to generalize the expressions for the transport coefficients in this situation (see for example Ref.~\cite{Bierkandt:2011zp}):
\be
\label{w-bulks}
\zeta_i (\omega) =\frac{1}{1+\left(\omega I^2_1 \,\frac{\partial \rho}{\partial n}\frac{\partial \rho}
{\partial \mu}\frac{T}{\Gamma_{ph}}\right)^2} \frac{T}{\Gamma_{ph}} \, C_i 
  \ , \qquad i=1,2,3,4 \ .
\ee
From the above expressions one can define the value of a characteristic frequency $\omega_c$ for the phonon collisions as 
\be
\omega_c = \frac{1}{I^2_1 \,\frac{\partial \rho}{\partial n}
\frac{\partial \rho} {\partial \mu}}
\frac{\Gamma_{ph}}{T} \  .
\label{char}
\ee
In the limit where $\omega \ll \omega_c$ one recovers the static bulk viscosity coefficients of Eqs.~(\ref{bulkstatic}).

After an expansion or rarefaction of the superfluid, the system  goes back to equilibrium after  a change in the number of phonons. Thus, the phonon decay rate $\Gamma_{ph}$ takes into account the first kinematically allowed number-changing processes. For superfluids where the phonons have a dispersion law with negative values of $\gamma$, the first kinematically allowed number-changing scattering consists of $2 \leftrightarrow 3$ collisions. The different $2 \leftrightarrow 3$ processes are given in Ref.~\cite{Manuel:2013bwa}.

\begin{figure}
\begin{tabular}{ccc}
\includegraphics[width=0.4\textwidth, height=5cm]{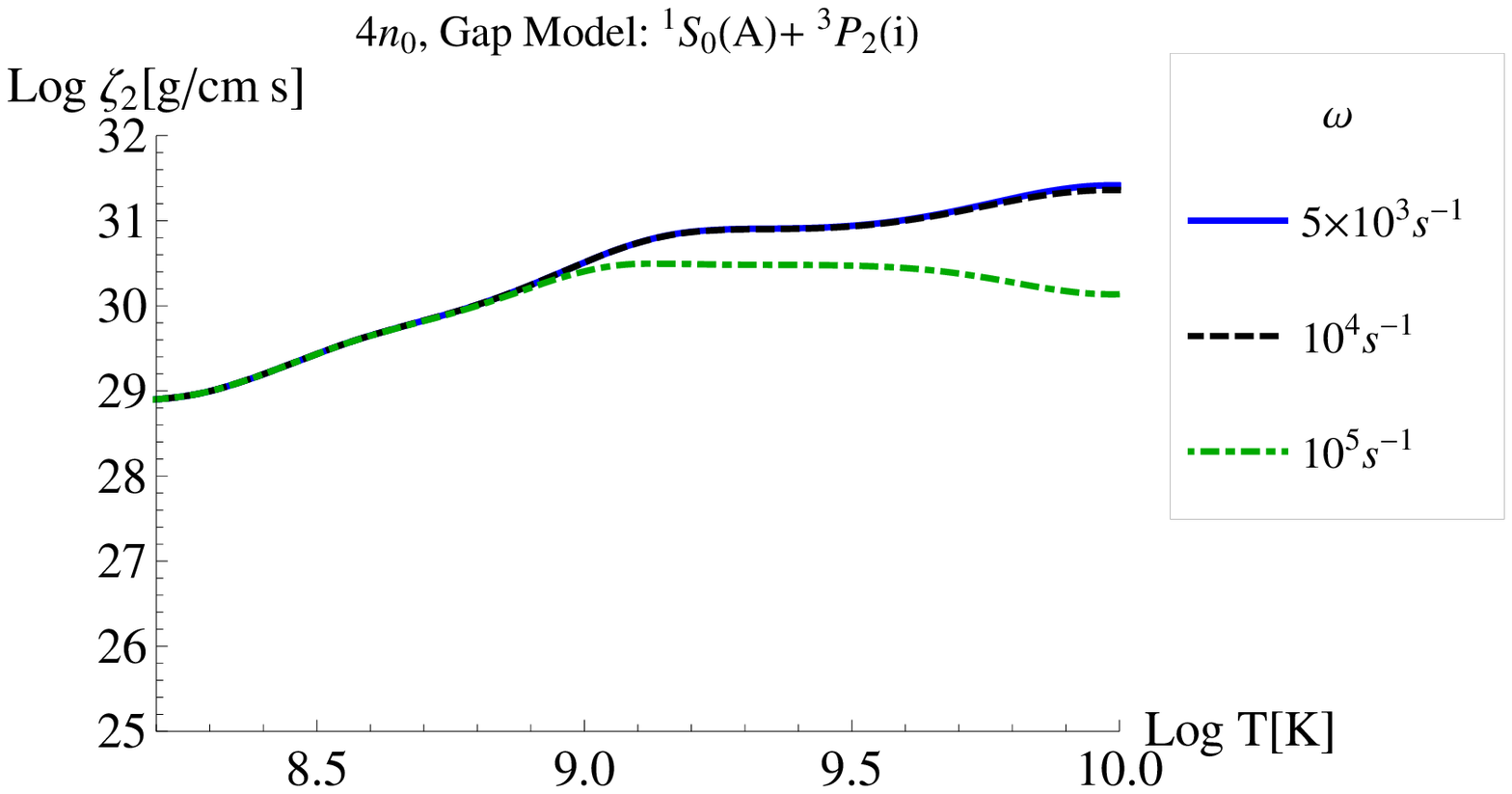} & \includegraphics[width=0.4\textwidth,height=5cm]{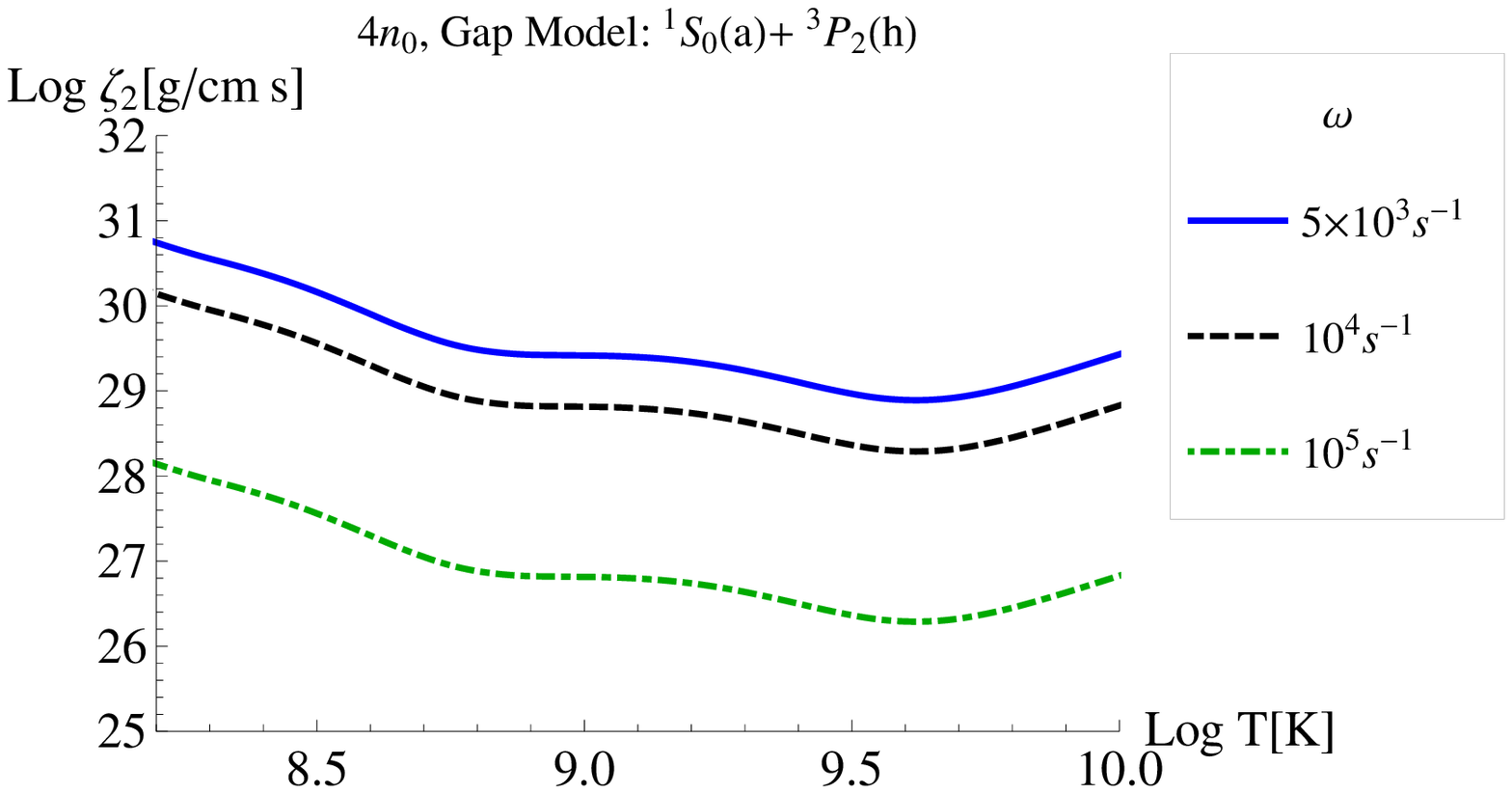}
\end{tabular}
\caption{ $\zeta_2$ frequency--dependent bulk viscosity coefficient as a function of the temperature for $4n_0$ and frequencies between $10^{3}-10^{5}s^{-1}$ for the two neutron gap models.}
\label{zeta2}
\end{figure}

In Fig.~\ref{zeta2} we display the frequency--dependent $\zeta_2$ coefficient  at $4n_0$, which describes the damping of stellar pulsations with typical frequencies of $\omega=10^3-10^5s^{-1}$.   When using the $^1S_0(A)+^3P_2(i)$ neutron gap model, we observe  on the l.h.s. of Fig.~\ref{zeta2} that for $\omega= 10^4s^{-1}$  the bulk viscosity is different by more than 10$\%$ from its static value only for $T\gtrsim10^{10}$K, while for $\omega= 10^5s^{-1}$ the difference is larger than 10$\%$ for $T\gtrsim10^9$K. On the contrary, the $\zeta_2$ coefficient is strongly dependent on the frequency if the  $^1S_0(a)+^3P_2(h)$ model is considered, as seen in the r.h.s. side of Fig.~\ref{zeta2}. Similar behavior is expected for the frequency--dependent $\zeta_1$ and $\zeta_3$ bulk viscosity coefficients. 

One should now compare our results for the frequency--dependent $\zeta_2$ bulk viscosity coefficient coming from the collisions among superfluid phonons with the contribution stemming from direct Urca \cite{Haensel:2000vz} and modified Urca \cite{Haensel:2001mw} processes for a typical  frequency of $\omega=10^4 s^{-1}$.   We find that,  at $T\sim 10^9$K and for typical radial pulsations of the star, phonon collisions give the leading contribution to the bulk viscosities in the core. We note that in Ref.~\cite{Manuel:2013bwa} a different conclusion was reached since the comparison was done taking into account Urca processes in normal matter. In that case, the phonon collisions give the leading contribution to the bulk viscosities in the core of the neutron stars, except when the opening of the Urca processes take place.

\section{Thermal conductivity}

In hydrodynamics the thermal conductivity   $\kappa$ is the coefficient that relates the heat flux with a temperature gradient as
 \ba
{\bf  q}=-\kappa \nabla T \ .
\label{defi}
 \ea
 
 We proceed similarly as in the shear viscosity case and we use variational methods in order to solve the transport equation. The final expression of the thermal conductivity reads  \cite{Manuel:2014kqa}
\ba
\kappa \geq \left(\frac{4a_1^2}{3 T^2}\right) A_1^2 M^{-1}_{11},
\label{kappa_var}
\ea
where $M^{-1}_{11}$ is the (1,1) element of the inverse of a matrix of $N \times N$ dimensions. Each element of this matrix is a multidimensional integral that contains
the thermally weighted scattering matrix for phonons. The number $N$ is treated as a variational parameter in our numerical study. 

It is interesting to note that for the computation of the superfluid phonon contribution to the thermal conductivity, the phonon dispersion law at NLO is required. This is so, as the thermal conductivity strictly vanishes if one uses a linear dispersion law \cite{Braby:2009dw}.

\begin{figure}
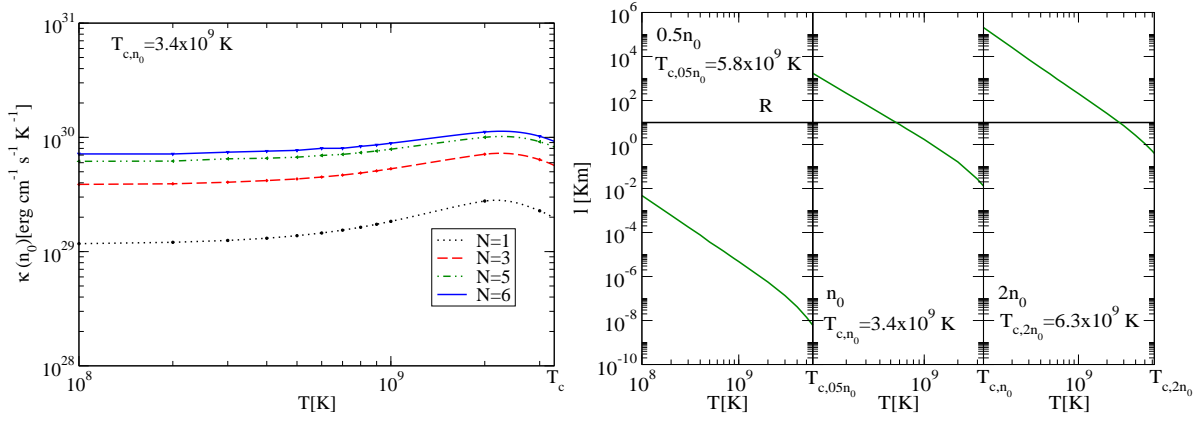

\includegraphics[width=0.45\textwidth]{therm_n_aip.eps}
\includegraphics[width=0.5\textwidth,height=5.5cm]{length_aip.eps}
 \caption{Left: Variational calculation for the thermal conductivity due to superfluid phonons up to order $N=6$ for  $n_0$ as a function of temperature. The calculation is performed  for the $^1S_0(A)$$+$$^3P_2(i)$ model for the gap. Right: mfp due to the phonon thermal conductivity using the $^1S_0(A)$$+$$^3P_2(i)$ model for the gap as a function of temperature for three different densities. The end temperature for each density is the critical temperature, $T_c=0.57 \Delta$. The mfp has to be compared with the radius of the star, which we take $R=10$ Km .}
 \label{fig:var-mfp}
\end{figure}

On the l.h.s of  Fig.~\ref{fig:var-mfp} we show the results up to order $N=6$ for  $n_0$ as a function of temperature. The final value of the number $N$ is determined by imposing that the deviation with respect to the previous order should be $\lesssim 10 \%$. The end temperature is the critical temperature which amounts for $T_c=0.57 \Delta(n_0)=3.4 \times 10^9$ K, with $\Delta(n)$ given on the r.h.s of Fig.~\ref{speeds-gap} for the $^1S_0(A)$$+$$^3P_2(i)$ model for the gap. We obtain that for T$ \lesssim 10^9$ K, below $T_c$, the thermal conductivity scales as $\kappa \propto 1/\Delta^6$, the factor of proportionality depending on the density.  Note that this is the same  temperature-independent behavior found for the color-flavor-locked superfluid
\cite{Braby:2009dw}. Close to $T_c$ we would expect that higher order corrections in the energy and momentum expansion should be taken into account both in the phonon dispersion law and self-interactions.

One can also extract the thermal conductivity mfp of the phonons, which is not the same as the mfp associated to shear viscosity (see Eqs.~(\ref{eq1},\ref{eq2}) for comparison). For the thermal conductivity 
the mfp  is defined as
\ba
l= \frac{\kappa}{\frac{1}{3} c_v c_s} ,
\ea
with the heat capacity for phonons given by \cite{IntroSupe}
\ba
c_v= \frac{2 \pi^2}{15 c_s^3} \left( T^3 + \frac{25 \gamma}{ 7} \frac{(2 \pi)^2}{c_s^2} T^5 \right) \ .
\ea

We show on the r.h.s. of Fig.~\ref{fig:var-mfp} the mfp of phonons in $\beta$-stable neutron star matter for $0.5 n_0$, $n_0$ and $2 n_0$ as a function of temperature using the $^1S_0(A)$$+$$^3P_2(i)$ model for the gap. 
We also indicate the limit of 10 Km for the radius of the star. We observe that  for $n= 0.5 n_0$ the superfluid phonon mfp stays below the radius of the star, also
 for $n= n_0$ and T $\gtrsim6\times10^8$ K  as well as for 2$n_0$ and T $\gtrsim3\times10^9$ K. We find that $l \propto 1/T^3$, resulting from the  temperature independent behaviour of the thermal conductivity. For the case of smaller gap values, such as for the  $^1S_0(a)$$+$$^3P_2(h)$, thermal conductivity can be orders of magnitude higher than the previous case and no hydrodynamical description is possible.

Our results should be compared to the thermal conductivity of electrons and muons mediated by electromagnetic interactions in neutron star cores. 
In Ref.~\cite{Shternin:2007ee} the calculations of the electron-muon contribution to the thermal conductivity have been revised.  We find that thermal conductivity in the neutron star core is dominated by phonon-phonon collisions when phonons  are in  a  hydrodynamical regime \cite{Manuel:2014kqa}. Our results also indicate that if the  contribution of electrons-muons and phonons to $\kappa$ become comparable, electron-phonon collisions could play an important role in the determination of $\kappa$. This deserves further studies.

\section{Summary}
Starting from a general formulation based on EFT for the collisions of superfluid phonons, we have computed the shear and bulk viscosities as well as the thermal conductivity due to phonons in the core of superfluid neutron stars. 

We have found that  the shear viscosity resulting from phonon binary collisions scales with $1/T^5$, a universal feature already seen for $^4$He and cold Fermi gases at unitary. The density dependence of shear viscosity is not universal, and depends on the EoS of the system.  We have also analyzed the effect of the shear viscosity due to phonons in the r-mode instability window and we have obtained that it  is modified for $T\gtrsim 10^8-10^9$K, the exact temperature depending on the star mass configuration.

We have also calculated the bulk viscosity coefficients due to phonons and concluded that they strongly depend on the gap of the superfluid neutron matter. However, phonon-phonon collisions seem to be the dominant process as compared to Urca and modified Urca processes for the determination of these viscous coefficients.

Finally we have found that the thermal conductivity due to phonons shows a T-independent behaviour for temperatures well below the critical one and it scales as $1/\Delta^6$, with the constant of proportionality depending on the EoS. The same behaviour with temperature was observed in the case of the color-flavor-locked phase. Moreover, phonon-phonon collisions dominate the thermal conductivity as compared to electron-muon collisions for typical temperatures inside the core of neutron stars.

Future work requires the careful analysis of electron-phonon collisions in the determination of these transport coefficients in the core of superfluid neutron stars.

\begin{theacknowledgments} This research was supported by Ministerio de Econom\'{\i}a y Competitividad under contracts FPA2010-16963 and FPA2013-43425-P, as well as NewCompStar (COST Action MP1304). LT acknowledges support from the Ramon y Cajal Research Programme from Ministerio de Econom\'{\i}a y Competitividad and from FP7-PEOPLE-2011-CIG under Contract No. PCIG09-GA-2011-291679.
\end{theacknowledgments}



\bibliographystyle{aipproc}   



\end{document}